\documentclass[aps,prb,twocolumn,showpacs,superscriptaddress]{revtex4-1}

\usepackage{amsmath}
\usepackage{graphicx}
\usepackage{color}
\usepackage{hyperref}
 
\begin{document}

\newcommand{\fab}[1]{{\bf \textcolor{red}{#1}}}


\title{Derivation of nonlinear single-particle equations \\
via many-body Lindblad superoperators: \\
A density-matrix approach}


\author{Roberto Rosati}
\affiliation{
Department of Applied Science and Technology, Politecnico di Torino \\
C.so Duca degli Abruzzi 24, 10129 Torino, Italy
}
\author{Rita Claudia Iotti}
\affiliation{
Department of Applied Science and Technology, Politecnico di Torino \\
C.so Duca degli Abruzzi 24, 10129 Torino, Italy
}
\author{Fabrizio Dolcini}
\affiliation{
Department of Applied Science and Technology, Politecnico di Torino \\
C.so Duca degli Abruzzi 24, 10129 Torino, Italy
}

\affiliation{CNR-SPIN, Monte S.Angelo - via Cinthia, I-80126 Napoli, Italy}
\author{Fausto Rossi}
\email[]{Fausto.Rossi@polito.it}
\homepage[]{staff.polito.it/Fausto.Rossi}
\affiliation{
Department of Applied Science and Technology, Politecnico di Torino \\
C.so Duca degli Abruzzi 24, 10129 Torino, Italy
}


\date{\today}

\begin{abstract}
A recently proposed Markov approach provides Lindblad-type scattering superoperators, which ensure the physical (positive-definite) character of the many-body density matrix. We apply the mean-field approximation to such many-body equation,  in the presence of one- and two-body scattering mechanisms, and we derive a closed equation of motion for the electronic single-particle density matrix, which turns out to be non-linear as well as non-Lindblad. 
We prove that, in spite of its nonlinear and non-Lindblad structure, the mean field approximation does preserve the positive-definite character of the single-particle density matrix, an essential prerequisite of any reliable kinetic treatment of semiconductor quantum devices.
This result is in striking contrast with conventional (non-Lindblad) Markov approaches, where the single-particle mean field equations can lead to positivity violations and to unphysical results. Furthermore, the proposed single-particle formulation is extended to the case of   quantum systems with spatial open boundaries, providing a formal derivation of a recently proposed density-matrix treatment based on a Lindblad-like system-reservoir scattering superoperator.
\end{abstract}

 

\maketitle

\section{Introduction}\label{s-I}

The microscopic derivation of suitable scattering superoperators is one of the most challenging problems in quantum physics. For purely atomic and/or photonic systems, dissipation and decoherence phenomena may successfully be described via adiabatic-decoupling procedures\cite{b-Scully97,b-Breuer07} in terms of extremely simplified models based on phenomenological parameters; within such  effective treatments, the main goal is to identify a suitable form of the Liouville superoperator, able to preserve the positive-definite character of the corresponding density-matrix operator.\cite{b-Davies76} This is usually accomplished by identifying proper Lindblad superoperators\cite{Lindblad76a} expressed in terms of a few crucial system-environment coupling parameters.
In contrast, in solid-state materials and devices the complex many-electron quantum evolution  results in a non-trivial interplay between coherent dynamics and energy-dissipation and decoherence processes,\cite{b-Bonitz98,b-Haug04,b-Datta05,b-Haug07,b-Jacoboni10,b-Rossi11} which has to be treated via fully microscopic approaches.

Based on the pioneering works by Van Hove,\cite{VanHove57a} Kohn and Luttinger,\cite{Kohn57a} and Zwanzig,\cite{Zwanzig61a} several adiabatic or Markov approximation schemes have been developed, which can be grouped in two main categories: approaches based on semiclassical (i.e., diagonal) scattering superoperators, also referred to as Pauli master equations,\cite{Fischetti99a,Gebauer04a,Knezevic08a} and fully quantum-mechanical (i.e., non-diagonal) dissipation models.\cite{Lindberg88a,Kuhn92a,Hohenester97a,BiSun99a,Flindt04a} These approaches have been widely applied to quantum-transport and coherent-optics phenomena in semiconductor materials and devices. 

When the system-environment coupling  becomes  strong and/or the excitation timescale is extremely short, markovian approaches are known to be unreliable, and memory effects have to be taken into account via quantum-kinetic approaches.~\cite{TranThoai93a,Schilp94a}
However, even in regimes where the Markov limit is applicable, conventional Markov approaches may lead to unphysical results. As originally pointed out by Spohn and co-workers.~\cite{Spohn80a} the positive-definite character of the density-matrix operator may be violated. In particular, in Ref.~[\onlinecite{Spohn80a}] the author pointed out that the choice of the adiabatic decoupling strategy is definitely not unique. Only the case discussed by Davies~\cite{b-Davies76} of a ``small'' subsystem   interacting with a thermal environment  could be shown to preserve positivity. However, such result was restricted to finite-dimensional subsystems (i.e., $N$-level atoms), and to the particular projection scheme of the partial trace. Thus, it can not be straightforwardly extended to the study of  solid-state systems.

To overcome this serious limitation, an alternative and more general Markov procedure has recently been proposed~\cite{Taj09b}, offering the following advantages: (i) in the discrete-spectrum case it coincides with the Davies model mentioned above, (ii) in the semiclassical limit it reduces to the well-known Fermi's golden rule, and (iii) it holds also in the case of a continuous-spectrum. Most importantly, this approach describes a genuine Lindblad evolution, thereby ensuring the positivity of the many-electron density matrix, and providing a reliable and robust treatment of energy-dissipation and decoherence processes in semiconductor quantum devices.
Once the evolution for the many-electron density matrix is thus solved, its positive-character --ensured by Lindblad evolution-- directly transfers via a trace-type projection onto the single-particle density matrix, which is needed to obtain various physical observables such as the carrier density, the average kinetic energy and the charge current. However, despite the conceptual importance of such alternative Markov approach, its practical implementation is limited by the fact that the many-body evolution is in general not exactly solvable, so that the single-particle density matrix cannot be extracted either. Indeed, so far, all relevant applications of such Markov treatment to semiconductor nanosystems are limited to the low-density limit,\cite{Dolcini13a,Rosati13b,Rosati14a,Rosati14b} where one can explicitly show that the scattering-induced time evolution is Lindblad-type also for the single-particle density matrix. 
At high carrier concentrations the problem is by far more complicated. A direct solution of the many-electron problem is in general too demanding, and approximations have to be introduced. The most straightforward way to obtain a closed equation for the single-particle density matrix is to apply the mean-field scheme to the many-electron equation. In doing so, however, the Lindblad structure of the original equation is lost. 
The crucial question arises whether the positive-character of the single-particle equation is violated by the mean field approximation.

Primary goal of this paper is to address this problem. By applying the conventional mean-field approximation to the many-electron dynamics obtained via the alternative Markov limit recalled above, we derive a closed equation of motion for the electronic single-particle density matrix, in the presence of one- as well as two-body scattering mechanisms. While  in the low-density limit the  Lindblad form is preserved, at finite or high carrier concentrations the equation turns out to be non-Lindblad and highly non-linear. Nevertheless, we are able to prove that the mean field approximation does preserve the positive-definite character of the single-particle density matrix, an essential prerequisite of any reliable kinetic treatment of semiconductor quantum devices.
Finally, the proposed single-particle formulation is  extended to the case of a quantum system with spatial open boundaries; this provides a formal derivation of a recently proposed density-matrix treatment based on a Lindblad-like system-reservoir scattering superoperator.\cite{Dolcini13a}

The Paper is organized as follows: In Sect.~\ref{s-mbsp}, after briefly recalling the main ingredients of the  alternative many-body Markov approach, we derive a nonlinear equation for the electronic single-particle density matrix in the presence of one- as well as two-body scattering mechanisms. Section \ref{s-pdc}  presents a detailed investigation, where we  show that the proposed single-particle equation preserves the positive-definite character of the single-particle density matrix. In Sect.~\ref{s-qsosb} our single-particle treatment is extended to the case of a quantum system with spatial open boundaries. Finally, in Sec.~\ref{s-SC} we summarize and draw our conclusions.

\section{From a many-body description to a single-particle picture}\label{s-mbsp}

Within the spirit of the usual perturbation theory, the global semiconductor Hamiltonian (electrons plus various crystal excitations, e.g., phonons, plasmons, etc.) may be written as 
\begin{equation}\label{Hglobal}
\hat{\mathbf{H}} = \hat{\mathbf{H}}_\circ + \sum_s \hat{\mathbf{H}}'_s\ ,
\end{equation}
where the first term $\hat{\mathbf{H}}_\circ$ is the unperturbed contribution that can be treated exactly, and the second term describes  a number of perturbations $\hat{\mathbf{H}}'_s$, corresponding to various interaction mechanisms (e.g., carrier-phonon, carrier-carrier, etc.), which are typically treated within some approximation scheme.\footnote{In what follows we shall limit ourselves to the case of time-independent interaction Hamiltonians; the generalization to time-dependent coupling mechanisms is straightforward.}

In the conventional approaches to the Markov limit, the second-order (or scattering) contribution to the time evolution of the global (e.g., carriers plus phonons) density-matrix operator $\hat{\boldsymbol{\rho}}$ can be written in operatorial form as
\begin{equation}\label{NLglobal}
\left.\frac{d \hat{\boldsymbol{\rho}}}{d t}\right|_{\rm scat}
= 
\frac{1}{2} \sum_s 
\left(
\hat{\mathbf{a}}^s \hat{\boldsymbol{\rho}} \hat{\mathbf{b}}^{s \dagger}
- 
\hat{\mathbf{a}}^{s \dagger} \hat{\mathbf{b}}^s \hat{\boldsymbol{\rho}}
\right)\ +\ {\rm H.c.}\ ,
\end{equation}
where $\hat{\mathbf{a}}^s = \frac{\hat{\mathbf{H}}'_s}{\hbar}$,
\begin{equation}\label{hatbfb}
\hat{\mathbf{b}}^s 
= 
\frac{1}{\hbar}
\int_{-\infty}^{+\infty} 
e^{-\frac{\hat{\mathbf{H}}_\circ t'}{i \hbar}}
\hat{\mathbf{H}}'_s
e^{\frac{\hat{\mathbf{H}}_\circ t'}{i \hbar}}
 dt'\ ,
\end{equation}
and H.c. denotes the Hermitian conjugate. The scattering superoperator in (\ref{NLglobal}) is definitely non-Lindblad, and therefore does not necessarily preserve the positive-definite character of the global density matrix $\hat{\boldsymbol{\rho}}$.~\cite{Iotti05a} 
  
In contrast, the alternative Markov procedure proposed in Ref.~[\onlinecite{Taj09b}] is based on a time symmetrization between microscopic and macroscopic scales, and enables one to express the scattering contribution 
in terms of the following Lindblad superoperator 
\begin{equation}\label{Lglobal}
\left.\frac{d \hat{\boldsymbol{\rho}}}{d t}\right|_{\rm scat}
= 
\sum_s 
\left(
\hat{\mathbf{A}}^s \hat{\boldsymbol{\rho}} \hat{\mathbf{A}}^{s \dagger}
- 
\frac{1}{2} \left\{\hat{\mathbf{A}}^{s \dagger} \hat{\mathbf{A}}^s , \hat{\boldsymbol{\rho}}\right\}
\right) \quad,
\end{equation}
where
\begin{equation}\label{hatbfA}
\hat{\mathbf{A}}^s 
= 
\lim_{\overline{\epsilon} \to 0}
\left(\frac{2 \overline{\epsilon}^2}{\pi\hbar^6}\right)^{1 \over
4} 
\int_{-\infty}^{+\infty}
e^{-\frac{\hat{\mathbf{H}}_\circ t'}{i \hbar}}
\hat{\mathbf{H}}'_s
e^{\frac{\hat{\mathbf{H}}_\circ t'}{i \hbar}}
 e^{-\left(\frac{\overline{\epsilon} t'}{\hbar}\right)^2} dt'\ .\end{equation}
with the energy $\overline{\epsilon}$ playing the role of the level broadening corresponding to a finite collision duration and/or to a finite single-particle life-time.\cite{b-Rossi11}

Both the non-Lindblad superoperator in (\ref{NLglobal}) and the Lindblad one in (\ref{Lglobal}) may suitably be expressed in terms of generalized scattering rates $\mathcal{P}^s$. More specifically, denoting by $\{\vert i \rangle\}$ and $\{\epsilon_i\}$ the eigenstates and the energy levels of the noninteracting Hamiltonian $\hat{\mathbf{H}}_\circ$, one obtains
\begin{equation}\label{DMglobal}
\left.\frac{d \boldsymbol{\rho}_{i_1i_2}}{d t}\right|_{\rm scat}
\!\!=\!\!
\frac{1}{2} \sum_{s,i'_1i'_2}
\left(\mathcal{P}^s_{i_1i_2,i'_1i'_2}
\!\boldsymbol{\rho}_{i'_1i'_2}
\!-\!
\mathcal{P}^{s *}_{i'_1i'_1,i_1i'_2}
\!\boldsymbol{\rho}_{i'_2i_2} \right)+{\rm H.c.} ,
\end{equation}
where for the non-Lindblad model in (\ref{NLglobal}) 
\begin{equation}\label{NLrates}
\mathcal{P}^s_{i_1i_2,i'_1i'_2} = \mathbf{a}^s_{i_1i'_1} \mathbf{b}^{s *}_{i_2i'_2}\ ,
\end{equation}
whereas for the Lindblad model in (\ref{Lglobal}) one has
\begin{equation}\label{Lrates}
\mathcal{P}^s_{i_1i_2,i'_1i'_2} = \mathbf{A}^s_{i_1i'_1} \mathbf{A}^{s *}_{i_2i'_2}\ .
\end{equation}
While for both models their diagonal (i.e., semiclassical) elements ($i_1i'_1 = i_2i'_2$) coincide with the standard Fermi's-golden-rule prescription,
\begin{equation}\label{FGR}
\mathcal{P}^s_{ii,i'i'} = \frac{2\pi}{\hbar} \left|\langle i \vert \hat{\mathbf{H}}'_s \vert i' \rangle\right|^2 \delta(\epsilon_i-\epsilon_{i'})\ ,
\end{equation}
the Lindblad form in (\ref{Lrates}) exhibits a more symmetric structure, a clear fingerprint of the time symmetrization previously mentioned.

The study of electro-optical processes in semiconductors mainly relies on physical quantities that depend on the electronic-subsystem coordinates only. It is thus customary to introduce a many-electron density-matrix operator  
\begin{equation}\label{trqp}
\hat{\boldsymbol{\rho}}_{\rm c} \doteq {\rm tr}\{\hat{\boldsymbol{\rho}}\}_{\rm p} \quad,
\end{equation}
where the non-relevant phononic (p) degrees of freedom have been traced out of the global density-matrix operator $\hat{\boldsymbol{\rho}}$.~\footnote{We stress that the proposed treatment of carrier-phonon interaction applies to other bosonic degrees of freedom as well (e.g., photons, plasmons, etc.).}
By denoting with $\hat{\boldsymbol{\rho}}^\circ_{\rm p}$ the equilibrium density-matrix operator of the phononic subsystem, and by assuming a state factorization of the form
\begin{equation}\label{cpfact}
\hat{\boldsymbol{\rho}} = \hat{\boldsymbol{\rho}}_{\rm c} \otimes \hat{\boldsymbol{\rho}}^\circ_{\rm p}\ ,
\end{equation}
it is possible to show\cite{Taj09b} that the reduced dynamics dictated by the Lindblad global evolution in (\ref{Lglobal}) is still of Lindblad type:
\begin{equation}\label{Lreduced}
\left.\frac{d \hat{\boldsymbol{\rho}}_{\rm c}}{d t}\right|_{\rm scat}
=
\sum_s \left(
\hat{\mathbf{A}}^s_{\rm c} \hat{\boldsymbol{\rho}}_{\rm c} \hat{\mathbf{A}}^{s \dagger}_{\rm c}
-
\frac{1}{2} \left\{\hat{\mathbf{A}}^{s \dagger}_{\rm c} \hat{\mathbf{A}}^s_{\rm c}, \hat{\boldsymbol{\rho}}_{\rm c}\right\}
\right) \ .
\end{equation}
Here the explicit form of the reduced or electronic operators $\hat{\mathbf{A}}^s_{\rm c}$ can be derived starting from the global Lindblad operators $\hat{\mathbf{A}}^s$ in (\ref{hatbfA}).\footnote{It is worth stressing that, in spite of their very same formal structure, Eqs.~(\ref{Lglobal}) and (\ref{Lreduced}) describe the system dynamics at different levels; this is confirmed by the fact that, while the global operators in (\ref{hatbfA}) are always Hermitian, the electronic operators $\hat{\mathbf{A}}^s_{{\rm c}}$ are usually non-Hermitian, a clear fingerprint of dissipation-versus-decoherence processes induced by the phononic subsystem on the carrier one.}

Within the above description, although a statistical average over the phononic degrees of freedom has been performed, the electronic subsystem is still treated via a many-body picture. Nevertheless, in the investigation of semiconductor-based quantum materials and devices, many of the physical quantities of interest are described via single-particle electronic operators of the form
\begin{equation}\label{hatbfG}
\hat{\mathbf{G}}_{\rm c} = \sum_{\alpha_1\alpha_2} G_{\alpha_1\alpha_2} \hat{c}^\dagger_{\alpha_1} \hat{c}^{ }_{\alpha_2}\ ,
\end{equation}
where $\hat{c}^\dagger_\alpha$ and $\hat{c}^{ }_\alpha$ denote the usual creation and destruction operators over the electronic single-particle states~$\vert\alpha\rangle$. 
Recalling that, for any electronic operator one has 
$
\langle \mathbf{G}_{\rm c} \rangle 
= {\rm tr}\{\hat{\boldsymbol{\rho}} \hat{\mathbf{G}}_{\rm c}\}
= {\rm tr}\{\hat{\boldsymbol{\rho}}_{\rm c} \hat{\mathbf{G}}_{\rm c}\}_{\rm c} 
$, 
the average value of the single-particle operator in (\ref{hatbfG})   can  be written as
\begin{equation}\label{av2}
\langle \mathbf{G}_{\rm c} \rangle = \sum_{\alpha_1\alpha_2} \rho_{\alpha_1\alpha_2} G_{\alpha_2\alpha_1}
\end{equation}
where 
\begin{equation}\label{SPDM}
\rho_{\alpha_1\alpha_2} = {\rm tr}\{ \hat{c}^\dagger_{\alpha_2} \hat{c}^{ }_{\alpha_1} \hat{\boldsymbol{\rho}}_{\rm c} \}_{\rm c}
\end{equation}
is the single-particle density matrix.

For the study of the time evolution of single-particle quantities, such as total carrier density, mean kinetic energy, charge current, and so on, it is then crucial to derive a closed equation of motion for the above single-particle density matrix. 
Combining its definition in (\ref{SPDM}) with the many-electron Lindblad dynamics in Eq.~(\ref{Lreduced}), and employing the cyclic property of the trace, one obtains
\begin{equation}\label{spe-gen}
\left.\frac{d \rho_{\alpha_1\alpha_2}}{d t}\right|_{\rm scat}
=
\frac{1}{2} \sum_s
{\rm tr}\left\{
\left[\hat{\mathbf{A}}^{s \dagger}_{\rm c}, \hat{c}^\dagger_{\alpha_2} \hat{c}^{ }_{\alpha_1} \right] \hat{\mathbf{A}}^s_{\rm c} \hat{\boldsymbol{\rho}}_{\rm c}
\right\}_{\rm c}\ +\ {\rm H.c.}\ .
\end{equation}

In order to get a closed equation of motion for the single-particle density matrix, it is now crucial to specify the form of our many-electron Lindblad operators $\hat{\mathbf{A}}^s_{\rm c}$ which, in turn, depends on the specific interaction mechanism considered. 

For the case of a generic carrier-phonon (cp) interaction mechanism the corresponding (one-body) Lindblad operator is always of the form
\begin{equation}\label{hatAcp}
\hat{\mathbf{A}}^s_{\rm c} 
= 
\sum_{\alpha\alpha'} A^{\rm cp}_{\alpha\alpha'} \hat{c}^\dagger_{\alpha} \hat{c}^{ }_{\alpha'}\quad.
\end{equation}
Equation (\ref{hatAcp}) describes the phonon-induced carrier transition  from the initial state $\alpha'$ to the final state $\alpha$. 
In this case the label $s \doteq {\bf q},\pm$  corresponds to the emission ($+$) or absorption ($-$) of a phonon with wavevector ${\bf q}$.

By inserting the carrier-phonon Lindblad operator (\ref{hatAcp}) into Eq.~(\ref{spe-gen}) and by employing the fermionic anticommutation relations, it is easy to show (see the Appendix for details) that the contribution to the system dynamics due to the generic carrier-phonon interaction mechanism~$s$ involves average values of four fermionic operators of the form:
\begin{equation}\label{h}
h_{\alpha_3\alpha_4,\alpha'_3\alpha'_4} = {\rm tr}\left\{
\hat{c}^\dagger_{\alpha_3} \hat{c}^{ }_{\alpha_4}
\hat{c}^\dagger_{\alpha'_3} \hat{c}^{ }_{\alpha'_4}
\hat{\boldsymbol{\rho}}_{\rm c} \right\}_{\rm c}\ .
\end{equation}

For the carrier-carrier (cc) interaction the Lindblad operator  has the general  two-body  form
\begin{equation}\label{hatAcc}
\hat{\mathbf{A}}^s_{\rm c} = 
\frac{1}{2} \sum_{\alpha\overline{\alpha},\alpha'\overline{\alpha}'} A^{\rm cc}_{\alpha\overline{\alpha},\alpha'\overline{\alpha}'} \hat{c}^\dagger_{\alpha} \hat{c}^\dagger_{\overline{\alpha}} \hat{c}^{ }_{\overline{\alpha}'} \hat{c}^{ }_{\alpha'} \quad,
\end{equation}
which describes the transition  of the electronic pair from the initial (two-body) state $\alpha'\overline{\alpha}'$ to the final state $\alpha\overline{\alpha}$.

As shown in Appendix, by inserting Eq.~(\ref{hatAcc}) in Eq.~(\ref{spe-gen}), the contribution to the system dynamics due to carrier-carrier interaction ($s = {\rm cc}$) involves average values of eight fermionic operators of the form:
\begin{equation}\label{k}
k_{\alpha_5\alpha_6\alpha_7\alpha_8,\alpha'_5\alpha'_6\alpha'_7\alpha'_8} \!=\! {\rm tr}\left\{
\hat{c}^\dagger_{\alpha_5} \hat{c}^\dagger_{\alpha_6}
\hat{c}^{ }_{\alpha_7} \hat{c}^{ }_{\alpha_8}
\hat{c}^\dagger_{\alpha'_5} \hat{c}^\dagger_{\alpha'_6}
\hat{c}^{ }_{\alpha'_7} \hat{c}^{ }_{\alpha'_8}
\hat{\boldsymbol{\rho}}_{\rm c} \right\}_{\rm c}\ .
\end{equation}

As anticipated, the crucial step in order to get a closed equation of motion for the single-particle density matrix consists of performing the well-known mean-field (or correlation-expansion) approximation;\cite{Axt98a,Rossi02b,Axt04a}
as discussed in Appendix, employing this approximation scheme and omitting renormalization terms,\cite{Rossi02b} for both carrier-phonon and carrier-carrier scattering the resulting single-particle equation is given by
\begin{widetext}
\begin{equation}\label{spe-cpcc}
\left.\frac{d \rho_{\alpha_1\alpha_2}}{d t}\right|_{\rm scat} = \frac{1}{2} \sum_{\alpha'\alpha'_1\alpha'_2} \left(\left(\delta_{\alpha_1\alpha'} - \rho_{\alpha_1\alpha'}\right)
\mathcal{P}^s_{\alpha'\alpha_2,\alpha'_1\alpha'_2} \rho_{\alpha'_1\alpha'_2} 
- 
\left(\delta_{\alpha'\alpha'_1} - \rho_{\alpha'\alpha'_1}\right) \mathcal{P}^{s *}_{\alpha'\alpha'_1,\alpha_1\alpha'_2} \rho_{\alpha'_2\alpha_2}\right)\ +\ {\rm H.c.}
\end{equation}
\end{widetext}
with generalized carrier-phonon scattering rates
\begin{equation}\label{calPcp}
\mathcal{P}^{s = {\rm cp}}_{\alpha_1\alpha_2,\alpha'_1\alpha'_2} = A^{\rm cp}_{\alpha_1\alpha'_1} A^{{\rm cp} *}_{\alpha_2\alpha'_2}
\end{equation}
and generalized carrier-carrier scattering rates
\begin{widetext}
\begin{equation}\label{calPcc}
\mathcal{P}^{s = {\rm cc}}_{\alpha_1\alpha_2,\alpha'_1\alpha'_2} = 2 \sum_{\overline{\alpha}_1\overline{\alpha}_2,\overline{\alpha}'_1\overline{\alpha}'_2} 
\left(\delta_{\overline{\alpha}_2\overline{\alpha}_1} - \rho_{\overline{\alpha}_2\overline{\alpha}_1}\right) 
\mathcal{A}^{\rm cc}_{\alpha_1\overline{\alpha}_1,\alpha'_1\overline{\alpha}'_1} 
\mathcal{A}^{{\rm cc} *}_{\alpha_2\overline{\alpha}_2,\alpha'_2\overline{\alpha}'_2} 
\rho_{\overline{\alpha}'_1\overline{\alpha}'_2} \ ,
\end{equation}
\end{widetext}
where
\begin{equation}\label{calA}
\mathcal{A}^{\rm cc}_{\alpha\overline{\alpha},\alpha'\overline{\alpha}'} 
\!=\! \frac{1}{4} \left(
A^{\rm cc}_{\alpha\overline{\alpha},\alpha'\overline{\alpha}'} 
\!-\!
A^{\rm cc}_{\overline{\alpha}\alpha,\alpha'\overline{\alpha}'} 
\!-\!
A^{\rm cc}_{\alpha\overline{\alpha},\overline{\alpha}'\alpha'} 
\!+\!
A^{\rm cc}_{\overline{\alpha}\alpha,\overline{\alpha}'\alpha'}
\right)
\end{equation} 
denote the totally antisymmetric parts of the two-body coefficients in (\ref{hatAcc}).

It is worth stressing that, differently from the generalized carrier-phonon rates in (\ref{calPcp}), the generalized carrier-carrier rates in (\ref{calPcc}) are themselves a function of the single-particle density matrix; this is a clear fingerprint of the two-body nature of the carrier-carrier interaction (see below).

The single-particle scattering superoperator in (\ref{spe-cpcc}) is the result of positive-like (in-scattering) and negative-like (out-scattering) contributions, which are nonlinear functions of the single-particle density matrix.
Indeed, in the semiclassical  limit,\cite{b-Rossi11} 
\begin{equation}\label{sl}
\rho_{\alpha_1\alpha_2} = f_{\alpha_1} \delta_{\alpha_1\alpha_2}\quad\ ,
\end{equation} 
the density-matrix equation (\ref{spe-cpcc}) assumes the expected nonlinear Boltzmann-type form
\begin{equation}\label{spe-cpcc-sl}
\left.\frac{d f_\alpha}{d t}\right|_{\rm scat} = \sum_{\alpha'} \left(
(1 - f_\alpha) P^s_{\alpha\alpha'} f_{\alpha'}  - (1 - f_{\alpha'}) P^s_{\alpha'\alpha} f_\alpha\right)
\end{equation}
with semiclassical carrier-phonon scattering rates
\begin{equation}\label{Pcp}
P^{s = {\rm cp}}_{\alpha\alpha'} = \mathcal{P}^{s = {\rm cp}}_{\alpha\alpha,\alpha'\alpha'} = \left\vert A^{\rm cp}_{\alpha\alpha'} \right\vert^2
\end{equation}
and semiclassical carrier-carrier scattering rates
\begin{equation}\label{Pcc}
P^{s = {\rm cc}}_{\alpha\alpha'} = \mathcal{P}^{s = {\rm cc}}_{\alpha\alpha,\alpha'\alpha'} = 
2 \sum_{\overline{\alpha}\overline{\alpha}'}
\left(1 - f_{\overline{\alpha}}\right) 
\left\vert \mathcal{A}^{\rm cc}_{\alpha\overline{\alpha},\alpha'\overline{\alpha}'}\right\vert^2 
f_{\overline{\alpha}'} \ .
\end{equation}

The above semiclassical limit clearly shows that the nonlinearity factors $(\delta_{\alpha_1\alpha_2} - \rho_{\alpha_1\alpha_2})$ in (\ref{spe-cpcc}) as well as in (\ref{calPcc}) can be regarded as the quantum-mechanical generalization of the Pauli factors $(1 - f_\alpha)$ of the conventional Boltzmann theory (see also Sec.~\ref{s-pdc}).

A closer inspection of Eqs.~(\ref{spe-cpcc}) and (\ref{calPcc}) ---together with their semiclassical counterparts in (\ref{spe-cpcc-sl}) and (\ref{Pcc})--- 
confirms the two-body nature of the carrier-carrier interaction. Indeed, differently from the carrier-phonon scattering, in this case the density-matrix equation describes the time evolution of a so-called ``main carrier''
$\alpha$ interacting with a so-called ``partner carrier'' $\overline{\alpha}$.

\section{Positivity analysis of the proposed single-particle equation}\label{s-pdc}

Primary goal of this section is to face the most important issue related to the proposed kinetic treatment: the positivity analysis of the nonlinear density-matrix equation in (\ref{spe-cpcc}). 
Indeed, if the single-particle density matrix describes a physical state, its eigenvalues are necessarily positive-definite and not greater than one (Pauli exclusion principle); in order to preserve such physical nature, it is imperative that the scattering-induced time evolution  preserves the values of the density-matrix eigenvalues within the interval [0,\ 1].

To this aim, let us start considering the case of carrier-phonon interaction previously discussed, whose nonlinear equation in (\ref{spe-cpcc}) (equipped with the generalized rates in (\ref{calPcp})) may also be easily rewritten in a more compact way via the one-electron operators 
\begin{equation}\label{hatrho}
\hat{\rho} = \sum_{\alpha_1\alpha_2} \vert \alpha_1 \rangle \rho_{\alpha_1\alpha_2} \langle \alpha_2 \vert
\end{equation}
and
\begin{equation}\label{hatA}
\hat{A}^{ } = \sum_{\alpha_1\alpha_2} \vert \alpha_1 \rangle A^{\rm cp}_{\alpha_1\alpha_2} \langle \alpha_2 \vert
\end{equation}
as
\begin{equation}\label{spe-cp-of}
\left.\frac{d \hat{\rho}}{d t}\right|_{\rm scat} 
= \frac{1}{2} \left(
(\hat{\mathcal{I}} - \hat{\rho}) \hat{A}^{ } \hat{\rho} \hat{A}^\dagger
- 
\hat{A}^\dagger (\hat{\mathcal{I}} - \hat{\rho}) \hat{A}^{ } \hat{\rho}\right)\ +\ {\rm H.c.}\ ,
\end{equation}
where $\hat{\mathcal{I}}$ denotes the identity operator of the one-electron Hilbert space.
Importantly, due to the quantum-mechanical Pauli factors $(\hat{\mathcal{I}} - \hat{\rho})$, the above scattering superoperator (\ref{spe-cp-of}) is nonlinear in $\hat{\rho}$ and non-Lindblad.\footnote{Indeed, at any time $t$ Eq.~(\ref{spe-cp-of}) can always be locally linearized\cite{b-Rossi11} treating the two Pauli factors $(\hat{\mathcal{I}} - \hat{\rho})$ as input parameters; however, the resulting (time-dependent) linear superoperator is definitely non-Lindblad.}
Only in the low-density limit, i.e. $\hat{\mathcal{I}} - \hat{\rho} \to \hat{\mathcal{I}}$, 
the nonlinear equation in (\ref{spe-cp-of}) reduces to the Lindblad superoperator
\begin{equation}\label{Lsp}
\left.\frac{d \hat{\rho}}{d t}\right|_{\rm scat} 
=
\hat{A}^{ } \hat{\rho} \hat{A}^\dagger
- 
\frac{1}{2} \left\{\hat{A}^\dagger \hat{A}^{ }, \hat{\rho}\right\}  \ ,
\end{equation}
and the positive-definite character of $\hat{\rho}$ is thereby ensured. At finite or high densities, no straightforward conclusion can be drawn about the positive-definite character of the generic time-dependent solution $\hat{\rho}(t)$.

Nevertheless, we show now that the proposed nonlinear single-particle equation in (\ref{spe-cpcc}) does preserve the positive-definite character of $\hat{\rho}$. In order to prove that,  let us describe the single-particle density matrix $\rho_{\alpha_1\alpha_2}$ via the corresponding operator $\hat{\rho}$ in (\ref{hatrho}); at any time $t$ it is possible to define its instantaneous (i.e., time-dependent) eigenvalues $\Lambda_\lambda$ and eigenvectors $\vert \lambda \rangle$ according to
\begin{equation}\label{eigen}
\hat{\rho} \vert \lambda \rangle = \Lambda_\lambda \vert \lambda \rangle\ ,
\end{equation}
which implies that
\begin{equation}\label{Lambda}
\Lambda_\lambda = \langle \lambda \vert \hat{\rho} \vert \lambda \rangle\ .
\end{equation}

The eigenvalues $\Lambda_\lambda$ in (\ref{eigen}) of a single-particle density matrix $\hat{\rho}$ describing a physical state are necessarily positive-definite and not greater than one (Pauli exclusion principle).
In order to preserve such positive-definite nature, it is imperative that the scattering-induced time evolution maintains the values of the eigenvalues within the physical interval [0,\ 1]; this can be verified by studying the time derivative of the generic eigenvalue in (\ref{Lambda}), namely:
\begin{equation}\label{dLambdadt}
\frac{d \Lambda_\lambda}{d t} =
\frac{d \langle \lambda \vert}{d t} \hat{\rho} \vert \lambda \rangle
+
\langle \lambda \vert \frac{d \hat{\rho}}{d t} \vert \lambda \rangle
+
\langle \lambda \vert \hat{\rho} \frac{d \vert \lambda \rangle}{d t}\ .
\end{equation}
In view of the completeness of the basis set $\{\vert \lambda \rangle\}$, the time derivative in (\ref{dLambdadt}) can also be written as:
\begin{eqnarray}\label{dLambdadtbis}
\frac{d \Lambda_\lambda}{d t} 
&=&
\sum_{\lambda'}
\frac{d \langle \lambda \vert}{d t} \vert \lambda' \rangle \langle \lambda' \vert \hat{\rho} \vert \lambda \rangle \nonumber \\
&+&
\langle \lambda \vert \frac{d \hat{\rho}}{d t} \vert \lambda \rangle \nonumber \\
&+&
\sum_{\lambda'} \langle \lambda \vert \hat{\rho} \vert \lambda' \rangle \langle \lambda' \vert \frac{d \vert \lambda \rangle}{d t}
\ .
\end{eqnarray}
Recalling that
\begin{equation}\label{rhome}
\langle \lambda \vert \hat{\rho} \vert \lambda' \rangle = \Lambda_\lambda \delta_{\lambda\lambda'}\ ,
\end{equation}
the result in (\ref{dLambdadtbis}) reduces to
\begin{equation}\label{dLambdadtter}
\frac{d \Lambda_\lambda}{d t} 
=
\Lambda_\lambda 
\frac{d \langle \lambda \vert}{d t} \vert \lambda \rangle 
+
\langle \lambda \vert \frac{d \hat{\rho}}{d t} \vert \lambda \rangle 
+
\Lambda_\lambda 
\langle \lambda \vert \frac{d \vert \lambda \rangle}{d t}\ .
\end{equation}
Taking into account that
\begin{equation}
\frac{d \langle \lambda \vert}{d t} \vert \lambda \rangle 
+
\langle \lambda \vert \frac{d \vert \lambda \rangle}{d t}
=
\frac{d \langle \lambda \vert \lambda \rangle}{d t} = 0\ ,
\end{equation}
the first and third term in (\ref{dLambdadtter}) cancel out exactly, and one finally concludes that
\begin{equation}\label{dLambdadtqua}
\frac{d \Lambda_\lambda}{d t} 
=
\langle \lambda \vert \frac{d \hat{\rho}}{d t} \vert \lambda \rangle
=
\frac{d \rho_{\lambda\lambda}}{d t}\ . 
 \end{equation}
This shows that the time variation of the eigenvalues $\Lambda_\lambda$ coincides with the time variation of the diagonal elements $\rho_{\lambda\lambda}$ of the operator $\hat{\rho}$ within the instantaneous eigenbasis $\{\vert \lambda \rangle\}$.

In order to evaluate the above time derivative, the crucial step is to analyze the explicit form of the proposed single-particle scattering superoperator written in the density-matrix eigenbasis of Eq.~(\ref{eigen}). Taking into account that the generic density-matrix equation (\ref{spe-cpcc}) is basis-independent, by replacing the original single-particle basis $\{\vert\alpha\rangle\}$ with the density-matrix eigenbasis $\{\vert\lambda\rangle\}$ and making use of Eq.(\ref{rhome}), its diagonal elements turn out to be:
\begin{equation}\label{spe-cp-lambda}
\frac{d \rho_{\lambda\lambda}}{d t} 
= 
\sum_{\lambda'}
\left[
(1-\Lambda_\lambda) P^s_{\lambda\lambda'} \Lambda_{\lambda'}
-
(1-\Lambda_{\lambda'}) P^s_{\lambda'\lambda} \Lambda_\lambda\right]\ ,
\end{equation}
where
\begin{equation}\label{Plambda}
P^s_{\lambda\lambda'} = \mathcal{P}^s_{\lambda\lambda,\lambda'\lambda'}
\end{equation}
are positive-definite quantities given by the diagonal elements of the generalized scattering rates (see Eqs.~(\ref{calPcp}) and (\ref{calPcc})) written in our instantaneous density-matrix eigenbasis.
By inserting this last result into Eq.~(\ref{dLambdadtqua}), one finally gets
\begin{equation}\label{BCT}
\frac{d \Lambda_\lambda}{d t}
=
\sum_{\lambda'}
\left[
(1-\Lambda_\lambda) P^s_{\lambda\lambda'} \Lambda_{\lambda'}
-
(1-\Lambda_{\lambda'}) P^s_{\lambda'\lambda} \Lambda_\lambda\right]\ .
\end{equation}
This last result is highly non-trivial: it states that, in spite of the partially coherent nature of the carrier dynamics in (\ref{spe-cpcc}), the time evolution of the eigenvalues $\Lambda_\lambda$ is governed by a non-linear Boltzmann-type equation, formally identical to the semiclassical result in (\ref{spe-cpcc-sl}).

We are now in the position to state that the physical interval [0,\ 1] is the only possible variation range of our eigenvalues $\Lambda_\lambda$. To this end, one can show that, when the latter approach the extremal values, $0$ or $1$, their time derivatives do not allow them to exit the interval.
Indeed, a closer inspection of the Boltzmann-like equation in (\ref{BCT}) shows that:
\begin{itemize}
\item[(i) ]
if one of the eigenvalues $\Lambda_\lambda$ is equal to zero, the corresponding time derivative in (\ref{BCT}) is always non-negative;
\item[(ii) ]
if one of the eigenvalues $\Lambda_\lambda$ is equal to one, its time derivative in (\ref{BCT}) is always non-positive.
\end{itemize}
\par\noindent
This leads us to the important conclusion that, for both  carrier-phonon and carrier-carrier scattering, the proposed nonlinear single-particle equation (\ref{spe-cpcc}) preserves the positive-definite character of the single-particle density matrix.

\begin{figure}
\centering
\includegraphics*[width=\columnwidth]{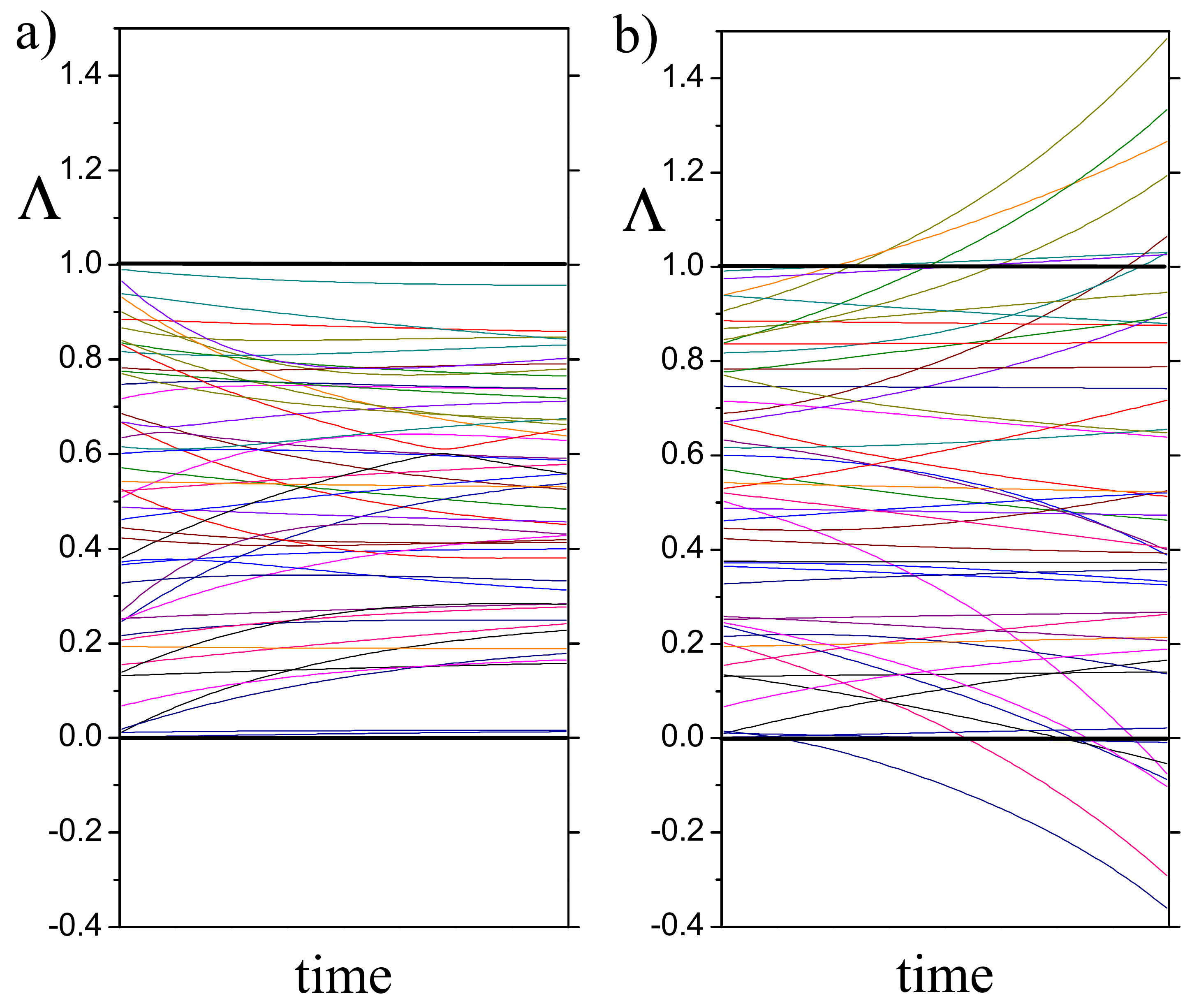}
\caption[]{(Color online)
Density-matrix eigenvalues as a function of time for a subset of 25 randomly generated evolutions corresponding to a simple two-level system in the presence of carrier-phonon interaction.
Comparison between the proposed single-particle model in (\ref{spe-cp-of}) (panel a) and the conventional model in (\ref{NLspe-cp-of}) (panel b) (see text).
}
\label{Fig1}       
\end{figure}

We finally stress that the above positivity analysis is based on the fact that the scattering rates in (\ref{Plambda}) are positive-definite quantities. This property, which applies to the proposed single-particle equation (obtained starting from the Lindblad-type scattering superoperator in (\ref{Lglobal})), is generally not fulfilled by conventional Markov models. In particular, for the case of carrier-phonon scattering, starting from the non-Lindblad scattering superoperator in (\ref{NLglobal}) and applying again the mean-field approximation, it is possible to derive a nonlinear single-particle equation of the form:
\begin{equation}\label{NLspe-cp-of}
\left.\frac{d \hat{\rho}}{d t}\right|_{\rm scat} 
= \frac{1}{2} \left(
(\hat{\mathcal{I}} - \hat{\rho}) \hat{a}^{ } \hat{\rho} \hat{b}^\dagger
- 
\hat{a}^\dagger (\hat{\mathcal{I}} - \hat{\rho}) \hat{b}^{ } \hat{\rho}\right)\ +\ {\rm H.c.}\ .
\end{equation}
This nonlinear equation is not intrinsically positive-definite, as confirmed by the fact that in the low-density limit the latter reduces to the following non-Lindblad form
\begin{equation}\label{NLsp-bis}
\left.\frac{d \hat{\rho}}{d t}\right|_{\rm scat} 
= \frac{1}{2} \left(
\hat{a}^{ } \hat{\rho} \hat{b}^\dagger
- 
\hat{a}^\dagger \hat{b}^{ } \hat{\rho}\right)\ +\ {\rm H.c.}\ .
\end{equation}
In analogy to Eq.~(\ref{NLrates}), the generalized scattering rates (within the eigenbasis $\{\vert\lambda\rangle\}$) corresponding to the above non-Lindblad superoperator are always of the form
\begin{equation}\label{calPlambda-NL}
\mathcal{P}^s_{\lambda_1\lambda_2,\lambda'_1\lambda'_2} = a^{ }_{\lambda_1\lambda'_1} b^*_{\lambda_2\lambda'_2}\ ,
\end{equation}
implying that in this case their diagonal elements
\begin{equation}\label{Plambda-NL}
P^s_{\lambda\lambda'} = \mathcal{P}^s_{\lambda\lambda,\lambda'\lambda'} = a^{ }_{\lambda\lambda'} b^*_{\lambda\lambda'}
\end{equation}
are not necessarily positive-definite.
This is the reason why, starting from a non-Lindblad many-body scattering model, the system dynamics may exit the physical eigenvalue region, giving rise to positivity violations also in the low-density limit.\cite{Iotti05a,Taj09b} 

To emphasize this point, in Fig.~\ref{Fig1} we report the time evolution of the density-matrix eigenvalues for a subset of simulated experiments in the simple case of a two-level system. As one can see, while for the proposed nonlinear equation in (\ref{spe-cp-of}) all the eigenvalue trajectories fall within the physical interval [0,\ 1] (see panel a)), for the nonlinear equation in (\ref{NLspe-cp-of}) a significant number of simulated eigenvalue trajectories exit the physical interval (panel b)). 
 
\section{Generalization to quantum systems with spatial open boundaries}\label{s-qsosb}

In what follows we extend the proposed single-particle treatment to quantum systems with spatially open boundaries, namely to the case of a quantum device electrically connected to one or more external carrier reservoirs.
To this end, in analogy to the system factorization between electronic and phononic degrees of freedom in (\ref{cpfact}), we shall describe the global carrier system (device plus reservoirs) as the product of a device density-matrix operator times a quasiequilibrium density-matrix operator corresponding to one or more carrier reservoirs:
\begin{equation}\label{drfact}
\hat{\boldsymbol{\rho}}_{\rm c} = \hat{\boldsymbol{\rho}}_{\rm d} \otimes \hat{\boldsymbol{\rho}}^\circ_{\rm r}\ .
\end{equation}

The quantum-mechanical coupling between device and external reservoirs ($s \equiv {\rm dr}$) may conveniently be described via the following interaction Hamiltonian
\begin{equation}\label{Hdr}
\hat{\mathbf{H}}'_s =
\sum_{\alpha\beta}
\left(
\gamma^{ }_{\alpha\beta} \hat{c}^\dagger_{\alpha} \hat{\xi}^{ }_{\beta}
+
\gamma^*_{\alpha\beta} \hat{\xi}^\dagger_{\beta} \hat{c}^{ }_{\alpha}
\right)\ ,
\end{equation}
where $\hat{c}^\dagger_\alpha$ ($\hat{c}^{ }_\alpha$) are now creation (destruction) operators acting on the device single-particle states $\alpha$, while $\hat{\xi}^\dagger_\beta$ ($\hat{\xi}^{ }_\beta$) denote creation (destruction) operators acting on the reservoir single-particle states $\beta$.
Here, the first contribution describes carrier injection ($\beta \to \alpha$) via the destruction of a carrier in state $\beta$ and the creation of a carrier in state $\alpha$, while the second one describes carrier loss ($\alpha \to \beta$) via the inverse process. 
Moreover, the physical properties of the device-reservoir interaction Hamiltonian in (\ref{Hdr}) are dictated by the explicit form of the coupling matrix elements $\gamma^{ }_{\alpha\beta}$; the latter, in general, are given by a properly weighted spatial overlap between device and reservoir single-particle wavefunctions.

Following the general prescription in (\ref{hatbfA}), the Lindblad operator corresponding to the device-reservoir interaction Hamiltonian (\ref{Hdr}) depends on the carrier coordinates only, and is always of the form
\begin{equation}\label{hatbfAdr}
\hat{\mathbf{A}}^s_{\rm c} =
\sum_{\alpha\beta}
\left(
A^{\rm dr}_{\alpha\beta} \hat{c}^\dagger_{\alpha} \hat{\xi}^{ }_{\beta}
+
A^{{\rm dr} *}_{\alpha\beta} \hat{\xi}^\dagger_{\beta} \hat{c}^{ }_{\alpha}
\right)\ .
\end{equation}

The evaluation of the single-particle dynamics induced by the above device-reservoir coupling may be performed following the very same steps of the corresponding carrier-phonon and carrier-carrier treatments previously considered and described in Appendix.
In particular, it is easy to realize that the single-particle contribution in Eq.~(\ref{spe-gen}) due to device-reservoir coupling involves average values of two device plus two reservoir creation/destruction operators.
More specifically, in view of the device-reservoir factorization in (\ref{drfact}) as well as of  the typical quasiequilibrium nature of the reservoirs, one obtains
\begin{equation}\label{avdr}
{\rm tr}\left\{
\hat{c}^\dagger_{\alpha_2} 
\hat{c}^{ }_{\alpha_1} \hat{\xi}^\dagger_{\beta_2} \hat{\xi}^{ }_{\beta_1}
\hat{\boldsymbol{\rho}}_{\rm c}\right\}_{\rm c}
=
\rho_{\alpha_1\alpha_2} \rho^\circ_{\beta_1\beta_2}\ , 
\end{equation}
where $\rho_{\alpha_1\alpha_2}$ is the single-particle density matrix of the device and
\begin{equation}\label{rhocirc}
\rho^\circ_{\beta_1\beta_2} = f^\circ_{\beta_1} \delta_{\beta_1\beta_2}
\end{equation}
is the (diagonal) single-particle density matrix of the quasiequilibrium carrier reservoirs.

Employing the device-reservoir factorization result in (\ref{avdr}), a straightforward calculation shows that the contribution to the system evolution due to the device-reservoir coupling Hamiltonian (\ref{Hdr}) is
\begin{equation}\label{spe-dr}
\left.\frac{d \rho_{\alpha_1\alpha_2}}{d t}\right|_{\rm scat} \!\!=\!\!
\frac{1}{2}\!
\sum_{\beta}
\!\left(\!
\mathcal{P}^s_{\alpha_1\alpha_2,\beta\beta}
\!f^\circ_{\beta} 
\!-\!
\sum_{\alpha'}
\mathcal{P}^s_{\alpha_1\alpha',\beta\beta}
\!\rho_{\alpha'\alpha_2}\!
\right)+{\rm H.c.} 
\end{equation}
with generalized scattering rates
\begin{equation}\label{calPdr}
\mathcal{P}^s_{\alpha_1\alpha_2,\beta_1\beta_2} = A^{\rm dr}_{\alpha_1\beta_1} A^{{\rm dr} *}_{\alpha_2\beta_2}\ .
\end{equation}

In the semiclassical limit (see Eq.~(\ref{sl})), the above device-reservoir scattering superoperator reduces to the relaxation-time model
\begin{equation}\label{spe-dr-sl-bis}
\left.\frac{d f_\alpha}{d t}\right|_{\rm scat} =
- \sum_{\beta}
P^s_{\alpha\beta}
\left(f_\alpha - f^\circ_\beta\right)
\end{equation}
with device-reservoir scattering rates
\begin{equation}\label{Pdr}
P^s_{\alpha\beta} = \mathcal{P}^s_{\alpha\alpha,\beta\beta} = \left\vert A^{\rm dr}_{\alpha\beta} \right\vert^2\ .
\end{equation}
It is worth stressing that the above semiclassical equation, usually referred to as the injection-loss model, has been widely employed in the semiclassical modeling of optoelectronic semiconductor devices.\cite{Iotti05b}

In order to gain more insight on the structure of the density-matrix equation (\ref{spe-dr}), the latter may conveniently be rewritten in a compact operatorial form; more specifically, recalling the definition of the single-particle density-matrix operator in (\ref{hatrho}) and introducing the device-reservoir coupling operators
\begin{equation}\label{hatAbeta}
\hat{A}_\beta = \sum_\alpha \vert\alpha\rangle A^{\rm dr}_{\alpha\beta} \langle\beta\vert
\end{equation}
as well as the reservoir density-matrix operator
\begin{equation}\label{hatrhocirc}
\hat{\rho}^\circ = \sum_\beta \vert \beta \rangle f^\circ_\beta \langle\beta\vert\ ,
\end{equation}
one gets:
\begin{equation}\label{Llike}
\left.\frac{d \hat{\rho}}{d t}\right|_{\rm scat}
=  
\sum_\beta \left(
\hat{A}^{ }_\beta \hat{\rho}^\circ \hat{A}^\dagger_\beta
- 
\frac{1}{2} \left\{\hat{A}^{ }_\beta \hat{A}^\dagger_\beta , \hat{\rho}\right\}
\right)\ .
\end{equation}  
Equation (\ref{Llike}) should be compared to the Lindblad superoperator in Eq.(\ref{Lsp}) describing energy exchange with the phononic excitations. On the one hand, the device-reservoir superoperator (\ref{Llike}) is inhomogeneous, due to the presence of the density-matrix operator $\hat{\rho}^\circ$ of the external reservoirs. This implies that the trace of the device density matrix $\hat{\rho}$ is not conserved, as expected in a system that can exchange particles with the reservoirs. On the other hand, the coupling term in (\ref{Llike}) is linear in $\hat{\rho}$ and has a Lindblad-like form, which ensures the positive-definite character of $\hat{\rho}$.  

The analysis presented so far can be regarded as a formal derivation of the Lindblad-like device-reservoir scattering superoperator recently proposed in Ref.~[\onlinecite{Dolcini13a}], where the reservoir states are plane waves ($\vert\beta\rangle = \vert k \rangle$) and the device single-particle states are the scattering states of the confinement potential profile ($\vert\alpha\rangle = \vert\alpha_k \rangle$).

\section{Summary and conclusions}\label{s-SC}

Exploiting a recent reformulation of the Markov limit, which enables one to provide genuine Lindblad-type scattering superoperators for the many-body density matrix, we have applied the mean-field approximation to the many-electron dynamics, and we have derived a closed equation of motion for the electronic single-particle density matrix, in the presence of carrier-phonon as well as carrier-carrier scattering mechanisms.  
While in the low-density limit the equation exhibits a Lindblad form ---like for the many-body density matrix--- at finite  carrier concentrations the resulting time evolution for the single-particle density matrix  turns out to be non-linear and non-Lindblad. 

We have proven (see Eq.~(\ref{BCT})) that, despite the lack of a Lindblad form,   the mean field approximation does preserve the positive-definite character of the single-particle density matrix, an essential prerequisite of any reliable and robust kinetic treatment of semiconductor quantum devices. This result is in striking contrast with the case of mean-field approximation applied to conventional (non-Lindblad) Markov approaches (see Eq.~(\ref{NLglobal})), where the corresponding single-particle equations may lead to positivity violations and thus to unphysical results.

The proposed single-particle formulation has then been extended to the case of quantum systems with spatial open boundaries; such microscopic treatment can be regarded as a formal derivation of a recently proposed density-matrix treatment\cite{Dolcini13a} based on a Lindblad-like system-reservoir scattering superoperator.

\acknowledgments
We are extremely grateful to David Taj for stimulating and fruitful discussions.
We gratefully acknowledge funding by the Graphene@PoliTo laboratory of the Politecnico di Torino, operating within the European FET-ICT Graphene Flagship project (www.graphene-flagship.eu).
Computational resources were provided by HPC@PoliTo, a project of Academic Computing of the Politecnico di Torino (www.hpc.polito.it). F.D. also  acknowledges financial support from Italian FIRB 2012 project HybridNanoDev (Grant No.RBFR1236VV).



\appendix*

\section{Derivation of the nonlinear single-particle scattering superoperator}\label{App}

In this appendix we recall the main steps involved in the derivation of the nonlinear single-particle equation in (\ref{spe-cpcc}).
To this aim, let us start by considering the case of carrier-phonon interaction. By inserting into Eq.~(\ref{spe-gen}) the carrier-phonon Lindblad operator in (\ref{hatAcp}) and employing the usual fermionic anticommutation relations, it is easy to show that
\begin{equation}
\left[ \hat{\mathbf{A}}^{s \dagger}_{\rm c}, \hat{c}^\dagger_{\alpha_2} \hat{c}^{ }_{\alpha_1} \right] 
= 
\sum_{\alpha'} 
\left(
A^{{\rm cp} *}_{\alpha_2\alpha'} \hat{c}^\dagger_{\alpha'} \hat{c}^{ }_{\alpha_1}
-
A^{{\rm cp} *}_{\alpha'\alpha_1} \hat{c}^\dagger_{\alpha_2} \hat{c}^{ }_{\alpha'}
\right)\ ,
\end{equation}
and therefore
\begin{eqnarray}\label{spe-cp}
\left.\frac{d \rho_{\alpha_1\alpha_2}}{d t}\right|_{\rm scat} &=&
\left(
\frac{1}{2}
\sum_{\alpha'\alpha'_1\alpha'_2}
A^{{\rm cp} *}_{\alpha_2\alpha'}
A^{\rm cp}_{\alpha'_1\alpha'_2}
h_{\alpha'\alpha_1,\alpha'_1\alpha'_2}
\right.\nonumber \\
&-& \left.\frac{1}{2}
\sum_{\alpha'\alpha'_1\alpha'_2} A^{{\rm cp} *}_{\alpha'\alpha_1}
A^{\rm cp}_{\alpha'_1\alpha'_2}
h_{\alpha_2\alpha',\alpha'_1\alpha'_2}
\right) + {\rm H.c.}\ ,
\nonumber \\
\end{eqnarray}
where $h$ is the two-particle correlation function introduced in Eq.~(\ref{h}).

The remaining step in order to get a closed equation of motion for the single-particle density matrix consists of performing the well-known mean-field (or correlation-expansion) approximation;\cite{Axt98a,Rossi02b,Axt04a} 
in this particular case this allows one to express the two-body correlation function (\ref{h}) ---given by the average value of four fermionic operators--- in terms of products of two single-particle density-matrix elements; more specifically, omitting renormalization terms,\cite{Rossi02b} one gets:
\begin{equation}\label{h-bis}
h_{\alpha_3\alpha_4,\alpha'_3\alpha'_4}
=
\left(\delta_{\alpha_4\alpha'_3} - \rho_{\alpha_4\alpha'_3}\right)
\rho_{\alpha'_4\alpha_3}\ .
\end{equation}
Employing this approximation scheme, the carrier-phonon scattering contribution in (\ref{spe-cp}) reduces to the single-particle density-matrix equation (\ref{spe-cpcc}) equipped with the generalized carrier-phonon rates in (\ref{calPcp}).

Let us finally come to the case of carrier-carrier interaction. 
In view of the usual anticommutation properties, the original carrier-carrier Lindblad operator in (\ref{hatAcc}) can also be written in terms of the fully antisymmetric coefficients in (\ref{calA}) as:
\begin{equation}\label{hatAcc-bis}
\hat{\mathbf{A}}^s_{\rm c} = 
\frac{1}{2} \sum_{\alpha\overline{\alpha},\alpha'\overline{\alpha}'} \mathcal{A}^{\rm cc}_{\alpha\overline{\alpha},\alpha'\overline{\alpha}'} \hat{c}^\dagger_{\alpha} \hat{c}^\dagger_{\overline{\alpha}} \hat{c}^{ }_{\overline{\alpha}'} \hat{c}^{ }_{\alpha'} \ .
\end{equation}
By inserting this alternative form of the Lindblad operator
into Eq.~(\ref{spe-gen}) and employing once again the usual fermionic anticommutation relations, it is easy to show that
\begin{eqnarray}
\left[ \hat{\mathbf{A}}^{s \dagger}_{\rm c}, \hat{c}^\dagger_{\alpha_2} \hat{c}^{ }_{\alpha_1} \right] 
&=& 
\sum_{\alpha\overline{\alpha}\alpha'} 
\mathcal{A}^{{\rm cc} *}_{\alpha'\alpha_2,\alpha\overline{\alpha}} 
\hat{c}^\dagger_{\alpha} \hat{c}^\dagger_{\overline{\alpha}} \hat{c}^{ }_{\alpha_1} \hat{c}^{ }_{\alpha'} 
\nonumber \\
&-&
\sum_{\overline{\alpha}\alpha'\overline{\alpha}'} 
\mathcal{A}^{{\rm cc} *}_{\alpha'\overline{\alpha}',\alpha_1\overline{\alpha}}
\hat{c}^\dagger_{\alpha_2} \hat{c}^\dagger_{\overline{\alpha}} \hat{c}^{ }_{\overline{\alpha}'} \hat{c}^{ }_{\alpha'} 
\nonumber \\
\end{eqnarray}
and therefore
\begin{widetext}
\begin{eqnarray}\label{spe-cc}
\left.\frac{d \rho_{\alpha_1\alpha_2}}{d t}\right|_{\rm scat} 
&=& 
\left( \frac{1}{4} 
\sum_{\alpha\overline{\alpha}\alpha',\alpha'_1\alpha'_2\alpha'_3\alpha'_4} 
\mathcal{A}^{{\rm cc} *}_{\alpha'\alpha_2,\alpha\overline{\alpha}} 
\mathcal{A}^{\rm cc}_{\alpha'_1\alpha'_2\alpha'_3\alpha'_4}
k_{\alpha\overline{\alpha}\alpha_1\alpha',\alpha'_1\alpha'_2\alpha'_4\alpha'_3}
\right.\nonumber \\
&-&
\left. \frac{1}{4} \sum_{\overline{\alpha}\alpha'\overline{\alpha}',\alpha'_1\alpha'_2\alpha'_3\alpha'_4} 
\mathcal{A}^{{\rm cc} *}_{\alpha'\overline{\alpha}',\alpha_1\overline{\alpha}}
\mathcal{A}^{\rm cc}_{\alpha'_1\alpha'_2\alpha'_3\alpha'_4}
k_{\alpha_2\overline{\alpha}\overline{\alpha}'\alpha',\alpha'_1\alpha'_2\alpha'_4\alpha'_3}
\right)\ +\ {\rm H.c.}\ ,
\end{eqnarray}
\end{widetext}
where $k$ is the four-particle correlation function introduced in Eq.~(\ref{k}).

Similarly to the case of carrier-phonon coupling, in order to get a closed equation of motion for the single-particle density matrix, one is forced to adopt the mean-field approximation scheme previously introduced. In this case, the latter amounts to writing the average values of eight fermionic operators in (\ref{k}) as products of four single-particle density-matrix elements. 
More specifically, neglecting again renormalization contributions, one gets:
\begin{widetext}
\begin{eqnarray}\label{k-bis}
k_{\alpha_5\alpha_6\alpha_7\alpha_8,\alpha'_5\alpha'_6\alpha'_7\alpha'_8} 
&=& 
\left(\delta_{\alpha_8 \alpha'_5}-\rho_{\alpha_8 \alpha'_5}\right) \left(\delta_{\alpha_7 \alpha'_6} - \rho_{\alpha_7 \alpha'_6}\right) \left(\rho_{\alpha'_7 \alpha_6} \rho_{\alpha'_8 \alpha_5 } - \rho_{\alpha'_8 \alpha_6} \rho_{\alpha'_7 \alpha_5}\right) \nonumber \\
&& -
\left(\delta_{\alpha_7 \alpha'_5}-\rho_{\alpha_7 \alpha'_5}\right) \left(\delta_{\alpha_8 \alpha'_6} - \rho_{\alpha_8 \alpha'_6}\right) \left(\rho_{\alpha'_7 \alpha_6} \rho_{\alpha'_8 \alpha_5 } - \rho_{\alpha'_8 \alpha_6} \rho_{\alpha'_7 \alpha_5}\right)\ .
\end{eqnarray}
\end{widetext}
Inserting the above mean-field factorization into Eq.~(\ref{spe-cc}),
after a straightforward calculation one gets again the nonlinear single-particle equation (\ref{spe-cpcc}), equipped with the generalized carrier-carrier rates in (\ref{calPcc}).


%

\end{document}